\documentclass[twocolumn,showpacs,preprintnumbers,amsmath,amssymb]{revtex4}

\usepackage{graphicx}
\usepackage{dcolumn}
\usepackage{bm}

\begin{document}


\title{Comment on
``Quasienergy anholonomy and its application to adiabatic quantum state manipulation''}

\author{Taksu Cheon}
  \affiliation{Laboratory of Physics, Kochi University of Technology,
    Tosa Yamada, Kochi 782-8502, Japan}

\date{November 1, 2007}

\maketitle

In their Letter \cite{TM07}, Tanaka and Miyamoto introduce a
kicked spin model, 
for which they point out the generic existence of exotic eigenvalue anholonomy.
They proceed to show the potential utility of the model for quantum state manipulation
and quantum information processing.
%

In our view, there are three missing elements in the article, without which
their discussion seems to be incomplete and the prospect of the application limited.
%

The first point is related to the different dynamical 
phase accumulated to different states during the variation of parameter
$\lambda$.  This would break the phase coherence of a superposed state which is
required to ensure quantum parallelism in quantum computation.  The cyclic
nature of the parameter variation in the model comes in for rescue, however.   
For two level systems, the difference of accumulated dynamical phase 
between two states 
$\left| 0 \right>$ and $\left| 1 \right>$ 
after the cyclic variation of $\lambda$ is
%
$
\int_0^{MT} (E_1-E_0) dt 
= \int_0^{2\pi} (E_1(\lambda)-E_0(\lambda)) 
{v_\lambda}^{-1}
d\lambda
$
%
where $v_\lambda=\frac{d \lambda}{d t}$ is the velocity of the variation 
of $\lambda$, $T$ the period of kick, $M$ the number of kicks 
to complete a $\lambda$-cycle. Since $v_\lambda$ is a quantity
at our disposal, which we may set as a constant $v_\lambda=2\pi/(MT)$
during the variation of $\lambda$, 
it could be used to ensure the phase difference becoming
an integer multiple of $2\pi$ so that the phase coherence between the two states is intact.  Possible modifications of above estimate by
step-by-step variation of  $\lambda$ will not change the essential 
story line that the dynamical phase difference is controllable by proper choice of $M$. 
%
%

The second point is on the proper consideration of Mead-Berry connection
hidden behind the scene.  Consider a {\it one-dimensionally mobile
kicked spin} described by the Hamiltonian
\begin{eqnarray}
\label{e02}
{\cal H} = \frac{1}{2M} P^2 
 + \frac{\pi}{2} \sigma_3 + \frac{R}{2}(1-\sigma_y) \sum_{m=-\infty}^{\infty} 
 \delta(t-mT)
\end{eqnarray}
with $P=-i\frac{d}{dR}$.  The second and the third term combined give exactly the 
original kicked spin model as appeared in FIG. 1 of \cite{TM07}
with $R$, the spatial coordinate variable, replacing the parameter $\lambda$ as the
coupling strength. First term represents the kinetic energy of the spin moving 
along the coordinate $R$.  
During a period $t=T$ (which we now set to be 1 for brevity),
the evolution is described by the unitary operator
\begin{eqnarray}
\label{e03}
U =e^{-i \frac{1}{2M} P^2}  [e^{-i\frac{\pi}{2}\sigma_3} e^{-i\frac{R}{2}(1-\sigma_y) }].
\end{eqnarray}
After diagonalizing the spin part of the system inside the bracket 
in the manner of \cite{TM07},
%
$
e^{-i\frac{\pi}{2}\sigma_3} e^{-i\frac{R}{2}(1-\sigma_y)} \phi_s(R) = e^{-iE_s(R)}\phi_s(R),
$
%
we write the eigenfunction $\Psi$  of the total Hamiltonian  ${\cal H}$ as
$\Psi = \sum_{s}\psi_s(R) \phi_s(R)$,
%
%
and obtain the quesieigenvalue equation 
\begin{eqnarray}
\label{e05}
\sum_{s'}\left[e^{-i\{\frac{1}{2M}({\bf P}-{\bf A})^2+{\bf E}\}}\right]_{ss'} \psi_{s'}(R) 
= e^{-i{\cal E}}\psi_s(R),
\end{eqnarray}
where ${\bf P}_{ss'}=-i\frac{d}{dR} \delta_{ss'}$, ${\bf E}_{ss'}=E_s(R) \delta_{ss'}$ are
momentum and effective potential matrices, respectively, and ${\cal E}$ the quasienergy
of the total Hamiltonina ${\cal H}$.
The gauge potential ${\bf A}$  is given by
\begin{eqnarray}
\label{e06}
{\bf A}_{ss'}(R) =  \left< \phi_s(R)\right| i\partial_R\phi_{s'}(R)\left.\!\right>,
\end{eqnarray}
which cannot be globally gauged out because of the eigenspace anholonomy. 
For the specific example of the model (\ref{e02}), 
it is given by ${\bf A}_{00}={\bf A}_{11}=0$, 
${\bf A}_{01}=-{\bf A}_{10}=-i/4$.  Diagonal elements of ${\bf A}$ are, 
in fact, always zero for kicked spin 1/2 with rank-1 perturbation,
guaranteeing the eigenspace anholonomy.  
Thus we have non-Abelian gauge structure here, just as in the
case of degenerate eigenvector anholonomy of Wilczek and Zee \cite{WZ84}. 

The above derivation of non-Abelian gauge leads to our third and final point.
In no place between (\ref{e03}) and (\ref{e06}), any assumption on the adiabaticity
of $R$ variable nor the concept of Born-Oppenheimer {\it approximation} invoked.
Equation (\ref{e05}) is exact, finite and well-defined as long as $s$ runs on finite dimension.

We conclude that, contrary to the view implied in \cite{TM07}, all the physics inherent
 in gauge potential obtained from degenerate Berry phase is replicated with models with eigenvalue anhonomy without any degeneracy,
 and that is achievable without the assumption of adiabaticity.

We acknowledge useful discussions with A.~Iqbal and K.~Miyazaki.  
This work is supported, in part, by
the Grant-in-Aid for Scientific Research of Japanese Ministry of Education
under the Grant number 18540384.

\end{document}